# Nano-imaging photoresponse in a moiré unit cell

Niels C.H. Hesp[1], Iacopo Torre[1], David Barcons-Ruiz[1], Hanan Herzig-Sheinfux[1], Kenji Watanabe[2], Takashi Taniguchi[3], Roshan Krishna Kumar[1]*, Frank H.L. Koppens[1,4]*

[1]ICFO-Institut de Ciencies Fotoniques, The Barcelona Institute of Science and Technology, 08860 Castelldefels (Barcelona), Spain.

[2]Research Center for Functional Materials, National Institute for Materials Science, Namiki 1-1, Tsukuba, Ibaraki 305-0044, Japan.

[3]International Center for Materials Nanoarchitectonics, National Institute for Materials Science, Namiki 1-1, Tsukuba, Ibaraki 305-0044, Japan.

[4]ICREA-Institució Catalana de Recerca i Estudis Avançats, 08010 Barcelona, Spain.

*To whom correspondence should be addressed: roshan.krishnakumar@icfo.eu, frank.koppens@icfo.eu

**Graphene-based moiré superlattices have recently emerged as a unique class of tuneable solid-state systems that exhibit significant optoelectronic activity. Local probing at length scales of the superlattice should provide deeper insight into the microscopic mechanisms of photoresponse and the exact role of the moiré lattice. Here, we employ a nanoscale probe to study photoresponse within a single moiré unit cell of minimally twisted bilayer graphene. Our measurements reveal a spatially rich photoresponse, whose sign and magnitude are governed by the fine structure of the moiré lattice and its orientation with respect to measurement contacts. This results in a strong directional effect and a striking spatial dependence of the gate-voltage response within the moiré domains. The spatial profile and carrier-density dependence of the measured photocurrent point towards a photo-thermoelectric induced response that is further corroborated by good agreement with numerical simulations. Our work shows sub-diffraction photocurrent spectroscopy is an exceptional tool for uncovering the optoelectronic properties of moiré superlattices.**

## Introduction

The photoresponse of semiconductor heterostructures is at the heart of modern optoelectronics[1,2]. A general prerequisite for photo-induced currents is the lack of an inversion centre, whether it is extrinsically defined by doping inhomogeneities in the form of PN junctions, or due to a broken inversion symmetry in the crystal structure. In this regard, the structures of moiré superlattices[3–7] are well suited for photoresponse applications, since the crystal symmetry can be easily reduced by a twist-angle ($\theta$) induced atomic-scale reconstruction[8,9]. Minimally twisted bilayer graphene (mTBG, $\theta < 0.1°$) represents such a class of moiré superlattices, in which lattice reconstruction of the twisted bilayer is energetically favourable and generates alternating triangular domains of AB/BA Bernal stacked regions separated by narrow domain wall networks[8,10,11] (Fig. 1a). Whereas the AB/BA regions locally are identical to Bernal stacked bilayer graphene, the atomic registry between AB sub lattices in each layer changes smoothly through the domain walls, dramatically changing the local electronic properties over length scales of $\sim 10$ nm[10]. From the perspective of photoresponse, those sharp changes in electronic spectra can serve as local junctions, thus providing an intrinsic photoactive region created by the moiré superlattice. Whilst these domain wall networks have already been shown to strongly influence the DC transport[12–15] and optical properties[16–23] of mTBG, their optoelectronic properties have not yet been explored. In part, this is challenging due to the sub-diffraction scale of the moiré that makes it difficult to resolve superlattice scale features in typical far-field photoresponse experiments[24,25].

In this work, we perform near-field photocurrent nanoscopy in mTBG. With this, we probe nanoscale photocurrents in mTBG and observe a unique photo-thermoelectric effect[26–29] governed by the symmetry breaking of the domain wall network. Our measurements are supported by simulations of the spatial photocurrent profile in our devices that shows the photoresponse originates from microscopic variations in the Seebeck coefficient intrinsic to the moiré lattice. By varying the doping, we observe anomalous sign reversals in the photocurrent that hints to the importance of strain gradients in mTBG. In addition, we observe a localised photoresponse close to the domain walls of the moiré unit cell that occurs by additional heating from polaritonic reflections.

**Results**

**Experimental techniques & device characterisation.** To study the photoresponse of our samples, we employed infrared scanning near-field photocurrent microscopy[30–32]. The technique involves local photo-excitation of carriers using a scattering-type scanning near-field optical microscope[33] (s-SNOM) combined with electrical current read out at one of the device contacts (see Methods). For local photoexcitation, we focus infrared light (with photon energy ∼100 – 200 meV) onto a sharp tip (Fig. 1a), inducing a strongly localised electric field at the tip apex in an area similar to the tip radius (20-30 nm). By scanning the tip around our sample, we build spatial maps of the photocurrent generated by this localized field. We note that in some cases we measured photovoltage, which is linearly related to the photocurrent through the device resistance $R$ as $I_{PC} = V_{PV}/R$. In this way, we can measure the photoresponse between two pairs of contact simultaneously. On the same sample, we measured the near-field optically scattered light from the tip using interferometric detection[34] (see Methods). This provides a complementary characterisation of the local optical response in our samples.

The studied mTBG samples were fabricated using the tear and stack method[7,35] (see Methods). In brief, the heterostructure structure consists of a Van der Waals stack of mTBG fully encapsulated with hexagonal boron-nitride[36] and deposited on top of a $SiO_2$/Si wafer. A sample (Fig. 1b) also contains gold contacts for photocurrent/photovoltage read out and simultaneous electrostatic gating of the channel with the Si backgate.

To characterise the local structure in our devices, we first measured the near-field optical scattering signal of our samples. Figure 1c shows a near-field image of the optically scattered light measured at high doping $n \sim 5 \times 10^{12}$ cm$^{-2}$ in our mTBG sample highlighted by the red box in Figure 1b (see Supplementary Section 1 for doping estimation). It reveals a set of bright fringes forming a triangular network. These features have already been studied extensively in near-field scattering experiments and were attributed to enhanced optical conductivity at the domain walls[20–22] of mTBG. Their observation in our experiment thus confirms the presence of atomic reconstruction[8] expected in mTBG and allows direct structural mapping of the moiré lattice in our samples.

**Photoresponse in mTBG.** Figure 1d shows a near-field photocurrent map of our sample measured in the mTBG region with an excitation energy $E$ = 188 meV, which exhibits a number of interesting features. First, a clear periodicity in the photocurrent is observed throughout the entire sample. It can be easily seen by following the zeros in photocurrent (white lines and white features) that trace the periodicity intrinsic to mTBG. Notably, the periodicity varies from ∼100 nm to ∼1000 nm in different regions of the device. We attribute this behaviour to local variations in the twist angle. Second, the photocurrent exhibits alternating domains of negative response (blue regions) and positive response (red regions). This is most clearly seen in the device region with largest moiré periodicity (black frame in Fig. 1d). Third, in these larger areas a second fringe can be seen running parallel to the Domain walls. This double-step like feature closely resembles that of a polaritonic reflection typically observed in near-field scattering and photocurrent experiments on graphene[31,37,38] and hBN crystal edges[39,40], and will be discussed in detail below. Importantly, this periodic structure in the photocurrent was characteristic in our other studied mTBG devices (see Supplementary Section 2).

The spatial pattern of photoresponse is highly sensitive to the position of measurement contacts with respect to the moiré periodicity. Figure 2a shows a zoomed photovoltage map of the marked area in Fig. 1d. This triangular pattern resembles the moiré pattern of mTBG (Fig. 1a), with zero crossings seemingly tracing the domain wall network of the moiré superlattice. However, when comparing our photoresponse data with our optical scattering data (Fig. 1c), we find that the actual structure of the moiré lattice is rather different. The black traces in Figure 2a outline the domain wall network of the moiré superlattice measured by optical scattering. From this we can see that this measurement does not capture the entire structure of the moiré lattice, and we lose a set of domain walls at 60º to those observed in Figure 2a. Strikingly, those additional domain walls appear simply by measuring between a different pair of contacts (Fig. 2b). In this scheme, the actual structure of the moiré lattice is clearly visible and the spatial profile of the photoresponse is particularly more complex than anticipated. For example, we find the measured photovoltage exhibits sign reversals not only at domain wall boundaries (Fig. 2c), but also within the AB/BA stacked domains. Moreover, the spatial patterns of the photoresponse within each moiré unit cell varies significantly between AB and BA stacking configurations.

To elucidate the role and mechanism of photoresponse in mTBG, we studied the gate-voltage dependence of the photocurrent within the moiré domains. In Figure 2d, we plot line traces of the measured photocurrent made across domain walls for different gate voltages. The positions (x) of domain walls are marked by the black arrows. The first thing to note, is that the gate-voltage dependence of the photoresponse is particularly sensitive to the position of the excitation spot within the moiré unit cell. On top of that, it is extremely non-monotonic, such that for some excitation positions it exhibits up to three sign changes. Figure 2e plots line cuts of the gate-voltage dependence for two excitation positions (marked by coloured dashed lines in Fig. 2d). On one hand, the observed non-monotonic behaviour strongly resembles that of the photo-thermoelectric effect in graphene[26,27,29] that is strongest for lower *n* and exhibits sign reversals at the charge neutrality point (CNP). This is not surprising considering the unit cell of mTBG is comprised mostly of AB Bernal stacked bilayer graphene, whose photoresponse is dominated by the photo-thermoelectric effect in the presence of spatially varying Seebeck coefficients[26,27,29]. On the other hand, some other features such as the two additional sign changes away from charge neutrality, which depend on the spatial location, are rather peculiar.

**Photo-thermoelectric effect.** With Figure 2 in mind, we constructed a model based on the photo-thermoelectric effect (PTE) to describe the observed photocurrent features in our device. Photocurrent generation due to PTE proceeds in four steps. First, the electric field generated by the tip induces an oscillating current density in the sample at the excitation frequency $\omega_{ph}$. Second, this oscillating current causes Joule heating of the electron gas. The power density produced is proportional to the square of the current and has a rectified component $Q(\boldsymbol{r}, \boldsymbol{r}_{tip})$, where $\boldsymbol{r}$ is the position and $\boldsymbol{r}_{tip}$ the position of the tip. Third, the generated heat spreads in the sample on a characteristic length scale referred to as the cooling length[28] ($L_{cool}$) before being dissipated to the substrate. Finally, since the heat transport is coupled to the charge transport via the Seebeck-Peltier effect, the heat flux is able to generate a net electric current in presence of Seebeck coefficient gradients. The current and heat flux are governed (at least for small incident power) by a set of linear equations with respect to the source term $Q(\boldsymbol{r}, \boldsymbol{r}_{tip})$. As a consequence, the PTE-induced photovoltage $V_{\text{PTE}}^{(m)}$ at the contact *m*, with respect to a grounded contact used as a common reference, is then given (see details in Supplementary Section 3) by a linear relation of the type

$$V_{\text{PTE}}^{(m)} = \int d\boldsymbol{r}\, \mathcal{R}_{\text{PTE}}^{(m)}(\boldsymbol{r})\, Q(\boldsymbol{r}, \boldsymbol{r}_{\text{tip}}), \tag{1}$$

where $\mathcal{R}_{\text{PTE}}^{(m)}(\boldsymbol{r})$ is the photovoltage responsivity function that encodes the PTE response of the system. An analogous formula holds for photocurrents. In absence of strong resonant features, for example in the case of plasmonic excitations, in the response of the system at the energy $\hbar\omega_{\text{ph}}$, $Q(\boldsymbol{r}, \boldsymbol{r}_{\text{tip}})$ rapidly decays with $|\boldsymbol{r} - \boldsymbol{r}_{\text{tip}}|$. This means that $V_{\text{PTE}}^{(m)} \propto \mathcal{R}_{\text{PTE}}^{(m)}(\boldsymbol{r}_{\text{tip}})$ and the photovoltage (or photocurrent) maps are essentially measuring the responsivity (apart from the convolution with a spread function due to the finite tip dimension).

To gain some intuition into the origin of the spatial profile of the photoresponse generated in our devices, we consider the form of the photovoltage responsivity function for a one-dimensional device (see derivation in Supplementary Section 4). This reads

$$\mathcal{R}(x) = -\frac{1}{\kappa W} \int dx' \frac{L_{\text{cool}}}{2} e^{-\frac{|x'-x|}{L_{\text{cool}}}} \partial_x S(x'), \tag{2}$$

where $W$ is the width of the device and $\kappa$ its electronic thermal conductivity. In this example (in open-circuit configuration) an infinitely sharp tip creates a spatial profile of electron temperature (Fig. 3a) in the form $\delta T(x) \equiv T(x) - T_0 \propto e^{-|x-x_{\text{tip}}|/L_{\text{cool}}}$, with $T_0$ being the temperature of the substrate. From (2) we clearly see that the PTE requires an inhomogeneous Seebeck coefficient with a gradient parallel to the path between measurement contacts. In addition to that, the responsivity decays away from the Seebeck coefficient fluctuations on a length scale $L_{\text{cool}}$, since that is the distance that heat is able to travel in the sample. Finally, in a system with particle-hole symmetry, $\mathcal{R}_{\text{PTE}}(x)$ is an odd function of the electronic density because $S(x)$ is odd. This behavior is clearly observed in the gate-voltage dependence of the measured photoresponse (Fig. 2e), where the photoresponse changes sign around the Dirac point due to a change in sign of the carriers' charge.

In the moiré superlattice of mTBG, the local layer alignment transitions smoothly in the region of the domain walls, which is expected to cause local gradients in the Seebeck coefficient, as required for the PTE. To corroborate this, we calculated (see Supplementary Section 5 and Methods for full details on the calculation) the Seebeck coefficient of bilayer graphene as a function of the layer alignment in the Relaxation Time Approximation[41] and mapped the local alignment to the distance from a domain wall using the result in Ref. 10. By this procedure we obtained the spatial profile of the Seebeck coefficient across a domain wall, as depicted in Figure 3b. This shows that the Seebeck coefficient dips sharply at the domain walls of the moiré superlattice. By considering equation (2) and recalling that $V_{\text{PTE}}^{(m)} \propto \mathcal{R}_{\text{PTE}}^{(m)}$, we plot the expected photovoltage profile in the vicinity of the domain walls (Fig. 3c). Notably, the photovoltage follows a non-monotonic dependence with position through the domain wall, changing sign as it crosses the saddle point. Such behaviour can indeed be seen in our data along certain domain walls (Fig. 2c).

Whilst the one-dimensional model (Fig. 3c) describes well the sign changes across certain domain walls (Fig. 2c), the full two-dimensional spatial map can be particularly more complex as one must consider the photocurrent contributions from different domain walls due to the finite cooling length $L_{\text{cool}}$. The full spatial profile of $\mathcal{R}_{\text{PTE}}^{(m)}(\boldsymbol{r})$ can be calculated if the entire measurement geometry of the system (Fig. 3d) and the spatial profile of the Seebeck coefficient (Fig. 3e) are known. We perform this calculation using the Finite Element Method (see Supplementary Section 3 and Methods for details). The calculation is greatly simplified thanks to a very elegant reciprocity relation[42,43].

Figure 3f and Fig. 3g plot the simulated photocurrent for the two measured contact configurations (Fig. 3d) in the same area corresponding to the measurement in Fig. 2a and Fig. 2b respectively. The simulations are carried out at a fixed doping $n = 1 \times 10^{12}$ cm$^{-2}$. Comparing Fig. 3f,g with Fig. 2a,b we find very good agreement between our measurements and the simulations. Not only do our simulations accurately capture the spatial sign changes observed in our sample, but also the differing local photoresponse between AB and BA stacked regions (Fig. 2b & Fig. 3g). Importantly, the simulations also capture the strong directional effect observed. To understand this behaviour, we recall that the PTE can generate a global current only in regions where the gradients of Seebeck coefficient contain a part running perpendicular to the projection of current flows between contacts. This means that the domains perpendicular to the projection of current flows contribute strongest to the measured photocurrent and are minimum for those that run parallel. The current projections for the two contact configurations are sketched in Fig. 3d, where the red and white lines depict projections for the contact configuration of Fig. 2a and Fig. 2b respectively. They are also drawn in Fig. 3e to illustrate their orientation relative to the domain wall networks measured in Fig. 2a,b. In the first configuration (red lines) we find that the domain walls that run almost parallel to the current projections. Therefore, photo-excitation at these domain walls does not contribute to the globally measured current, and they remain completely hidden in our measurement (Fig. 2a). Whereas in the second case (white lines), there is always a component perpendicular to all sets of domain walls such that they all contribute to the globally measured current (Fig. 2b). This result also demonstrates the crucial importance of measurement geometry[44] in understanding the nanoscale photoresponse of solid-state crystals.

Good agreement between the simulations and experiment Fig. 2a,b points towards a photo-thermoelectric dominated photoresponse in the moiré lattice of mTBG. Indeed, qualitatively the spatial sign changes across domain walls and doping dependent sign changes around the charge neutrality point are well described by our photo-thermoelectric model. Even the additional sign changes observed away from CNP (see purple trace in Fig. 2e), can be explained within the framework of the PTE by considering an enhanced Seebeck coefficient at the domain walls to that which was calculated (see Supplementary Section 6). That said, it does not accurately describe the spatial dependence of the gate-voltage response, specifically, the sign reversal of the photoresponse within the moiré domains shifting with applied gate voltage (dashed line in Fig. 2d). This behaviour points towards a spatially varying parameter not considered in our model, for example, a spatially varying Seebeck coefficient within the AB/BA domains themselves that is not localised at the domain walls. Such behaviour might be expected in the case that lattice reconstruction in mTBG imposes significant strain in the AB regions[45], which might enhance the Seebeck coefficient of bilayer graphene locally[46]. A simplified model, which includes a spatially varying Seebeck within the AB/BA domains, is presented in Supplementary Section 6 and shows similar features as the experimental data.

Another peculiarity in the data is the high doping behaviour ($|V_G - V_D| > 40$ V), in which the spatial sign changes across domain walls becomes nearly absent (Fig. 2d) and the spatial profile resembles that of a constant background, which changes sign when changing carrier polarity. This is illustrated in Fig. 4a that plots a photocurrent map for a doping $n \sim 4 \times 10^{12}$ cm$^{-2}$, where we observe large areas of the moiré lattice exhibiting either a constant positive or negative photoresponse. In the simplest case, we might attribute this additional photocurrent to nearby PN junctions caused by deformations/stacking faults in our heterostructure. However, our measurements of the cooling length from these interfaces ($L_{cool}$ = 240 nm) show a fast decay of such contributions (see Supplementary Section 7). Hence, the data suggests another photocurrent mechanism might be present in mTBG. For example, we considered the possibility of photogalvanic currents (see Supplementary Section 8). Further work is required to understand these additional background phenomena.

**Photoresponse from polaritonic reflections.** Finally, we address the double step-like feature that is observed close to the domain walls (Fig. 1d, Fig. 2d, Fig. 4a). At first inspection, those features resemble that of polariton reflections from crystal edges that are typically observed in near-field scattering/photocurrents experiments in graphene-hBN heterostructures[31,37–40,47,48]. Notably, the domain walls of mTBG behave somewhat like crystal edges and have been shown to reflect propagating plasmon polaritons[20,21] and even phonon polaritons in encapsulating hBN proximity layers[49]. As shown in this work, the domain walls also act as local photoactive junctions which would thus enable thermoelectric detection of polariton modes[31,40]. Indeed, the double-step feature measured at 117 meV (Fig. 4a) resembles the plasmonic reflections previously observed in scattering-type SNOM experiments on domain walls in bilayer graphene[20,21,49]. For investigation, we studied the wavelength dependence of the double-step feature focussing on energies 150 - 200 meV where the double-step feature appeared strongest (Fig. 1d). Figure 4b plots the photoresponse as a function of excitation wavelength for a line trace made across several domain walls in one of our mTBG samples. For all energies, we observe the expected photocurrent profile generated at domain walls by the photo-thermoelectric effect (Fig. 3c). However, we also observe an additional feature that disperses with energy in the specific range 180 – 200 meV but is completely absent for lower energies. Considering this energy range corresponds to the upper Reststrahlen band of hBN, those dispersing features resemble that of phonon-polaritons scattering from hBN crystal edges. For further analysis, Figure 4c compares line traces of the measured photoresponse (green arrow in Fig. 4b) at different energies after subtraction of a smooth background; the background is experimentally extracted from the line trace at 202 meV where the double-step feature is not observed. The traces clearly show the peaks shifting in position away from the domain wall with lowering excitation energy. Assuming these as polariton reflections[40,49], we can extract the wavelength $\lambda_p$ and plot corresponding energy dispersion for these excitations. Figure 4d plots the experimentally measured dispersion (red circles) with theoretical calculations[39,50,51] of the loss function of hBN phonon polaritons. The good agreement provides strong evidence this feature is caused by hBN phonon-polaritons and shows the importance of polariton reflections in the photoresponse of our samples (Fig. 1d).

**Discussion**

To conclude, we show that near-field photocurrent spectroscopy is a valuable tool for studying the optoelectronic properties of moiré superlattices. Our moiré-scale resolved measurements reveal a spatially rich photoresponse governed by the symmetries of the reconstructed lattice that would go unnoticed in typical far-field photoresponse experiments. Good agreement of our simulations with experimental data shows the importance of hot carriers in the photoresponse of mTBG and, at the same time, shows the crucial link between global measurements and local excitation in photocurrent experiments. Our work should thus motivate further near-field photocurrent studies on related moiré superlattices including twisted transition metal dichalcogenides[9] and small-angle twisted bilayer graphene.

## Methods

### Device fabrication

The mTBG device discussed in the main text was fabricated using a standard tear-and-stack method[7,35] with a set twist angle of 0.15°. Here we use a polycarbonate film to pick up the individual flakes to form a hBN(30nm)/mTBG/hBN(6nm), which is released at 180 °C on a Si/SiO$_2$(285 nm) substrate. 1D contacts are patterned by photolithography, followed by Cr(5 nm)/Au(50 nm) deposition[36]. To avoid picking up contamination during our measurements, we clean the surface from polymers residues by the AFM-brooming[52]. Other samples were fabricated in a similar manner.

### Measurement details

We used an s-SNOM from Neaspec that allows simultaneous measurement of the scattered near-field signal and any photocurrent/photovoltage signals. We used a CO2 laser (Access Laser) and a tuneable QCL laser (Daylight Solutions) as infrared light sources between 110-250 meV (5-11 μm), at a typical power between 20-40 mW. This light is focused on a PtIr-coated AFM tip (Nanoworld), which is oscillating above the sample surface at ~250 kHz with a tapping amplitude of 80-100 nm. The modulated scattering signal is detected by a fast cryogenic HgCdTe detector (Kolmar Technologies), and by operating the s-SNOM in pseudoheterodyne mode we can independently measure the scattering amplitude and phase[34]. For photocurrent measurements we used a fast current amplifier (Femto DLPCA-100). For simultaneous measurement of the photovoltage between two pairs of contacts, we used two differential voltage amplifiers (Ithaco 1201) with one common contact grounded. The carrier doping in our samples is tuned by applying a DC voltage between the Si backgate and our device. To avoid detecting unwanted far-field contributions to the scattered or photocurrent/voltage signal, we detect the near-field signals at the 2$^{nd}$ or 3$^{rd}$ harmonic of the cantilever oscillation.

The measured photocurrent/voltage signal is demodulated with the driving signal of the AFM cantilever as reference signal. However, the actual motion of the AFM cantilever can have a phase offset that varies with the position on the sample (due to tip-sample interaction). This phase offset is given at each pixel by the measured phase delay between the tip driving signal and the actually detected motion. To correct our photocurrent/voltage signal measured at harmonic $i$, we subtract at every point $i$ times this phase delay. In addition to this, there remains a global phase offset in the corrected photocurrent/voltage signal due to the electronics in the circuit. Since the photocurrent/voltage signal is a real-valued quantity, we subtract this global phase offset, which we determine by taking the most frequent phase within a scan.

### Numerical simulations

Finite Element Method (FEM) simulations of the PTE response of the device were performed using an open-source (LGPLv3), home-made, python package (available at https://gitlab.com/itorre/diffusive_solver)[43] based on the FEniCS library[53], that allows the solution of coupled diffusion equations systems in realistic sample geometries. Band structure calculations of shifted graphene bilayers and the extraction of the corresponding physical properties were carried out using an open-source (LGPLv3) python package (available at https://gitlab.com/itorre/bandstructure-calculation)[54], which allows the computation of the spectrum and of optical and thermoelectric properties of simple electronic band-structure models.

**Acknowledgements**

We thank Justin Song for fruitful discussions, and are grateful to Matteo Ceccanti for providing the illustration in Fig. 1a.

Research leading to these results has received funding from the European Union's Horizon 2020 research and innovation programme under grant agreement Ref. 881603 (Graphene flagship Core3). Furthermore, this work was supported by the ERC TOPONANOP under grant agreement Ref. 726001.

N.C.H.H. also acknowledges funding from the European Union's Horizon 2020 programme under the Marie Skłodowska-Curie grant agreement Ref. 665884. I.T. also acknowledges funding from the Spanish Ministry of Science, Innovation and Universities (MCIU) and State Research Agency (AEI) via the Juan de la Cierva fellowship Ref. FJC2018-037098-I. D.B.R. also acknowledges funding from the Secretaria d'Universitats i Recerca del Departament d'Empresa i Coneixement de la Generalitat de Catalunya, as well as from the European Social Fund. H.H.S. also acknowledges funding from the European Union's Horizon 2020 programme under the Marie Skłodowska-Curie grant agreement Ref. 843830. K.W. and T.T. also acknowledge support from Japan's MEXT under the Elemental Strategy Initiative (grant Ref. JPMXP0112101001), from a KAKENHI grant by the JSPS (Ref. JP20H00354), and from the JST under the CREST programme (grant Ref. JPMJCR15F3). R.K.K. also acknowledges funding from the European Union's Horizon 2020 programme under the Marie Skłodowska-Curie grant agreements with Ref. 754510 and Ref. 893030. F.H.L.K. also acknowledges support from the Government of Spain (FIS2016-81044; Severo Ochoa CEX2019-000910-S), Fundació Cellex, Fundació Mir-Puig, and Generalitat de Catalunya (CERCA, AGAUR, SGR 1656).

**Author contributions**

F.H.L.K and N.C.H.H. conceived the experiment. N.C.H.H. and R.K.K. conducted the experiments, with assistance from D.B.R. and H.H.S.. Devices were fabricated by N.C.H.H. and R.K.K., with K.W. and T.T. providing hBN crystals. N.C.H.H., I.T., R.K.K. and F.H.L.K. analysed and interpreted the results. N.C.H.H. and I.T. performed numerical simulations using the band structure calculator and FEM solver developed by I.T.. The manuscript was written by R.K.K., I.T., N.C.H.H, and F.H.L.K. with input from all authors. F.H.L.K. supervised the work.

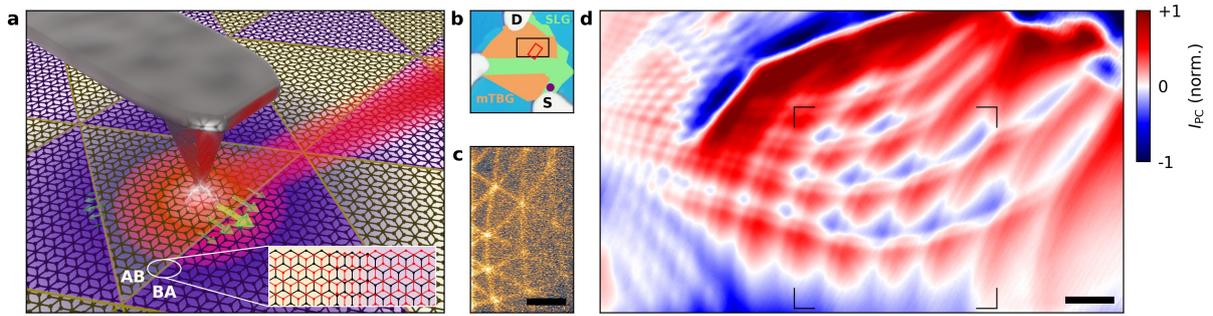

**Figure 1 Near-field photoresponse in minimally twisted bilayer graphene. a** Schematic illustration of near-field photocurrent experiments performed in small-angle twisted bilayer graphene. The moiré domains of different stacking configurations are highlighted by yellow and purple shaded areas. The AFM tip couples the infrared light into the device, causing the electron temperature to elevate locally. The photo-thermoelectric effect converts this heat partially in a current that can be read out by the contacts. Inset: Zoom of the domain wall structure separating AB-BA domains of the more lattice. **b** Device schematic of the main device under study. It consists of different regions of single layer graphene (SLG) and mTBG due to folding/stacking faults during the heterostructure assembly. The geometry for photocurrent measurements in panel **d** is marked by S (source) and D (Drain). The third contact is left floating. The purple dot marks the position where the calibration of carrier density is done. **c** Near-field scattering phase image of our sample in the mTBG region, measured at $E$ = 117 meV and $n \sim 5 \times 10^{12}$ cm$^{-2}$. Length of scale bar: 500 nm. **d** Photocurrent map of one of our mTBG samples measured at carrier density $n \sim 1 \times 10^{12}$ cm$^{-2}$ with an excitation energy $E$ = 188 meV. Length of scale bar: 500 nm. The map is normalised by the maximum measured $I_{PC}$.

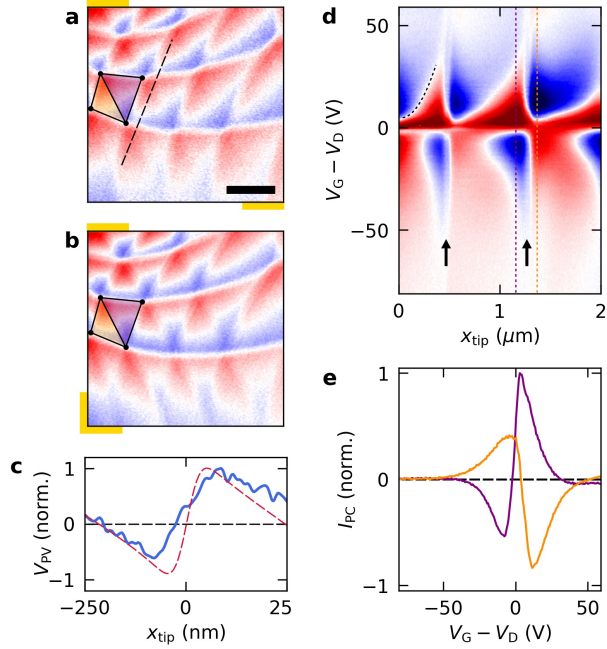

**Figure 2 Photoresponse in the moiré unit cell of mTBG. a** Zoomed map of measured photovoltage $V_{PV}$ in the same sample presented in Fig. 1 (corresponding to the marked area in Fig. 1**d**), measured at excitation energy $E$ = 117 meV and carrier density $n \sim 1 \times 10^{12}$ cm$^{-2}$. The map is normalised by the maximum measured $V_{PV}$. Colour code: blue: -1, red: +1. The gold annotations illustrate roughly the relative position of contacts used for measuring voltages. **b** same as panel **a** but for a different choice of contacts (gold annotations) measured simultaneously. **c** Line-cut across a horizontal domain wall in panel **b** (black), together with a line trace of the simulated profile in Fig. 3**g** (dashed red). **d** Photocurrent as a function of gate voltage $V_G$ (with respect to the position of the Dirac point $V_D$) measured for a line trace that crosses several domain walls (see dashed line in Fig. 2**a**). Black arrows mark the position of domain walls (except for those that contribute weakly to the photocurrent in this orientation). The gate voltages correspond to carrier densities within roughly $\pm 6 \times 10^{12}$ cm$^{-2}$. **e** Gate-voltage response along the two line-cuts highlighted in panel **d**, on two opposite sides of the domain wall.

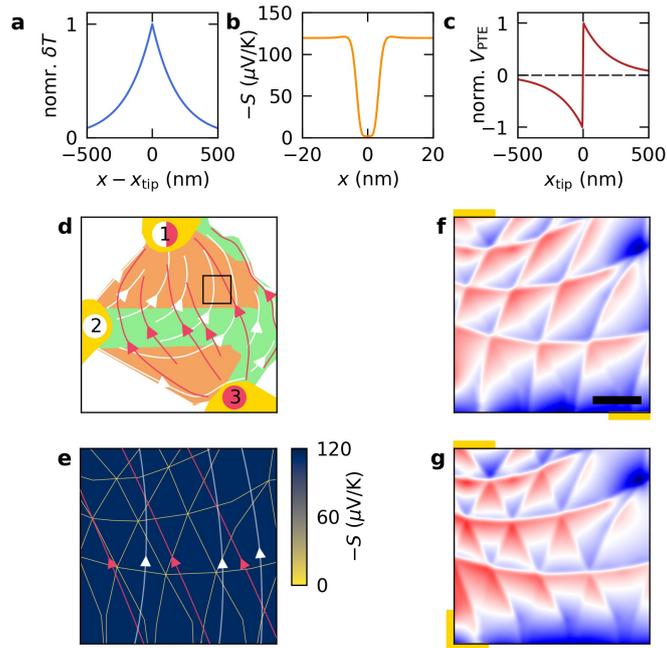

**Figure 3 Simulations of the photo-thermoelectric response in mTBG. a** Profile of the electron temperature increase induced by local photoexcitation as a function of distance ($x$) from the excitation position ($x_{tip}$), for a cooling length of 200 nm. **b** Calculated Seebeck coefficient as a function of position within two moiré unit cells separated by a domain wall at $x$ = 0 nm. **c** Line trace of the calculated photovoltage across a domain wall using the thermal profile and Seebeck coefficient of panel **a** and **b**. **d** Illustration of the geometry used in our photocurrent simulations. The yellow semi-circles indicate contact probes, the green single layer graphene regions and the orange mTBG regions. The red projection of current flows corresponds to a configuration wherein current flows between contacts 3 and 1, while the yellow field lines correspond to current flowing between contacts 2 and 1. The black square highlights the region shown in panels **e**-**g**. **e** 2D spatial map of the calculated Seebeck coefficient (see Supplementary Section 5) that goes into our model, along with the projections of current flows shown in panel **d**. **f** Zoom of the photovoltage simulations in the region of mTBG (same area as Fig. 2**a**). Here we convolute the calculated responsivity with a Gaussian function of width 15 nm to account for the finite tip radius. **g** Same as in panel **f** but for a different measurement geometry (same as Fig. 2**c**).

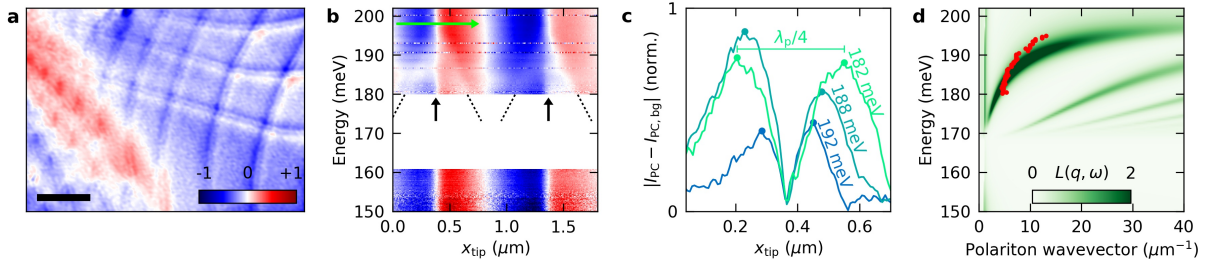

**Figure 4 Anomalous photocurrent and polariton reflections. a** Spatial map of the normalised photocurrent ($I_{PC}$) at an excitation energy of 117 meV and doping $n \sim 4 \times 10^{12}$ cm$^{-2}$. Length of scale bar: 500 nm. **b** Normalised $I_{PC}$ as a function of excitation energy and excitation position ($x_{tip}$) crossing two domain walls (marked by black arrows) in another of our mTBG samples. The dashed lines are guides to the eye showing the dispersive nature of the double-step feature. The white area corresponds to a gap in the spectrum of our laser. **c** Line traces (as absolute value) of a few energies taken from panel **b** (see green arrow in panel **b**) plotted after background subtraction. The distance between the peaks corresponds to a quarter of the polariton wavelength $\lambda_p$. **d** Calculated loss function of phonon-polaritons in hBN (see Ref. 51 for details). The red circles plot the experimentally extracted dispersion of the observed features.

**Supplementary Information**

# Nano-imaging photoresponse in a moiré unit cell

Niels C.H. Hesp[1], Iacopo Torre[1], David Barcons-Ruiz[1], Hanan Herzig-Sheinfux[1], Kenji Watanabe[2], Takashi Taniguchi[3], Roshan Krishna Kumar[1]*, Frank H.L. Koppens[1,4]*

[1]ICFO-Institut de Ciencies Fotoniques, The Barcelona Institute of Science and Technology, 08860 Castelldefels (Barcelona), Spain.

[2]Research Center for Functional Materials, National Institute for Materials Science, Namiki 1-1, Tsukuba, Ibaraki 305-0044, Japan.

[3]International Center for Materials Nanoarchitectonics, National Institute for Materials Science, Namiki 1-1, Tsukuba, Ibaraki 305-0044, Japan.

[4]ICREA-Institució Catalana de Recerca i Estudis Avançats, 08010 Barcelona, Spain.

*To whom correspondence should be addressed: roshan.krishnakumar@icfo.eu, frank.koppens@icfo.eu

This file contains the following Supplementary Sections:

1. **Carrier density calibration**
2. **Photoresponse in other mTBG devices**
3. **Photo-thermoelectric effect**
4. **Photo-thermoelectric effect in a 1D channel**
5. **Calculation of thermoelectric transport coefficients in mTBG**
6. **Impact of strain inside the AB domains**
7. **Cooling length in our devices**
8. **Effects beyond the photo-thermoelectric effect**

**Supplementary references**

## Supplementary Section 1: Carrier density calibration

In field-effect transistor geometries, the carrier density ($n$) induced by an applied gate voltage is generally well described by a simple capacitance model. Even so, Hall-effect measurements are usually a pre-requisite in proper characterisation of $n$ in any conducting system and allows one to make a calibration of the $n$ induced by the applied gate voltage ($V_G$). In our devices, however, measurement geometries are not-well suited for Hall measurements. To accurately determine the induced $n$ in our devices, we instead perform a calibration by measuring the plasmon-polariton dispersion in a single-layer graphene (SLG) region. Infrared plasmon-polaritons have been studied extensively in graphene, and their dispersions are well known[1–3]. By studying the near-field photocurrent close to a PN junction, we directly image graphene plasmons, measure their wavelength, and determine the doping level $n$ that such an excitation corresponds to. Moreover, tuning the gate-voltage tunes the plasmon wavelength, which allows us to make a calibration of $n(V_G)$. Figure S1a plots a map of the measured photovoltage (we plot the derivative with respect to $x_{tip}$ to make features clearer) as a function of gate voltage and tip position ($x_{tip}$), where the x-axis corresponds to a spatial line scan made near one of the Au measurement contacts of our device (purple dot in Fig. 1b of the main text). The Au-SLG interface is marked by the dashed line in Fig. S1a. In line with the photo-thermoelectric effect, the photovoltage changes sign when the gate is tuned through the charge neutrality point of graphene. On top of this, we can also see a set of fringes that become wider spaced at higher gate voltages. They arise from thermoelectric detection[2,3] of the interfering plasmon-polaritons in graphene. To extract the plasmon wavelengths, we follow the method described in Ref. 1, that involves fitting a polynomial combined with a sinusoidal function (see inset of Fig. S1b). The corresponding $n$ for the measured plasmon wavelength is then determined from the plasmon dispersion relation (Fig. S1b) calculated in our sample at the excitation energy used in our measurement (117 meV)[1]. With this method, we obtain the density calibration plotted in Fig. S1c. It shows a linear behaviour with gate-voltage as expected. Note, $n$ is slightly higher than what is typically expected for the dielectric thickness of our capacitor, which we attribute to photodoping[4] caused by the constant far-field infrared illumination that is unavoidable in our SNOM experiments.

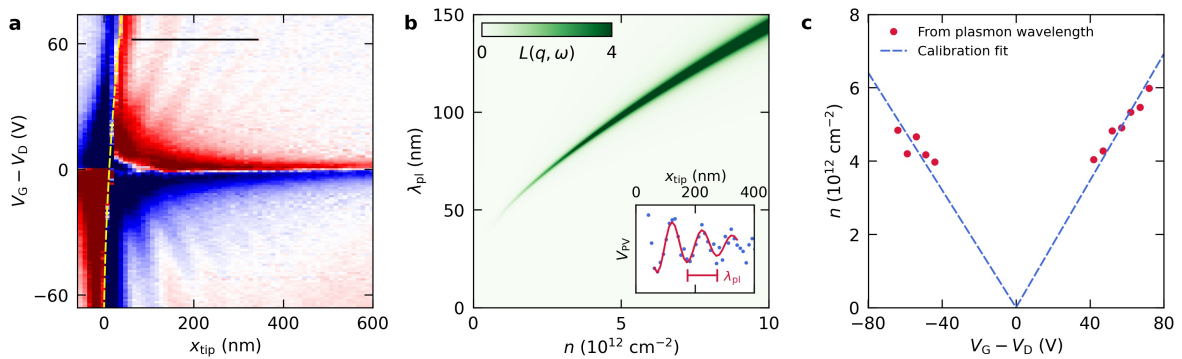

**Figure S1 Estimation of the induced carrier density for an applied gate voltage. a** Derivative along $x_{tip}$ of the measured photovoltage in single-layer graphene near a contact for a range of gate voltages. The contact (extending up to the yellow dashed line) serves as a launcher for plasmon-polaritons in SLG, observed as periodic oscillations in the plot. The excitation energy is 117 meV. **b** Calculated loss function of SLG in our sample at an excitation energy of 117 meV for various carrier densities. The sharp green line corresponds to the plasmon-polariton resonance, which changes its wavelength $\lambda_{pl}$ with carrier density. Inset: measured photovoltage along the black trace in panel **a**, together with a fit (red line) to extract $\lambda_{pl}$. **c** Extracted carrier density for various gate voltages based on the data of panel **a** and the calculated dispersion relation shown in panel **b**. The blue dashed lines are linear fits on either side of the CNP, giving an estimation of the induced carrier density at lower gate voltages.

**Supplementary Section 2: Photoresponse in other mTBG devices**

The behaviour reported in the main text was found to be generic to our other measured mTBG devices. To illustrate this, we plot measured photocurrent maps of another mTBG device (Fig. S2a). Although slightly more anisotropic, the triangular patterns of the moiré lattice intrinsic to mTBG can be seen and are similar to those measured in the contact configuration used in Fig. 2a of the main text. We find the same qualitative behaviour including sign changes across certain domain walls, in line with what is expected from the PTE (Fig. 3a-c of the main text), and the double-step feature at domain wall interfaces (see dotted lines in Fig. S2a).

The measurements presented in the main text were focused on mTBG with moiré structures of size ~500 nm. However, we also observed structures with smaller periodicities. Figure S2b plots the measured photocurrent of the device presented in the main text, but in the mTBG region on the bottom side of the single-layer graphene region. Again, we find the same periodic patterns in the photocurrent as observed in larger structures. Whilst the seemingly square lattice we observe in Fig. S2b is not representative of the moiré lattice in mTBG, we recover the triangular lattice (see shaded domains) by incorporating the same directional effect as to that reported in the main text (Fig. 2a of the main text).

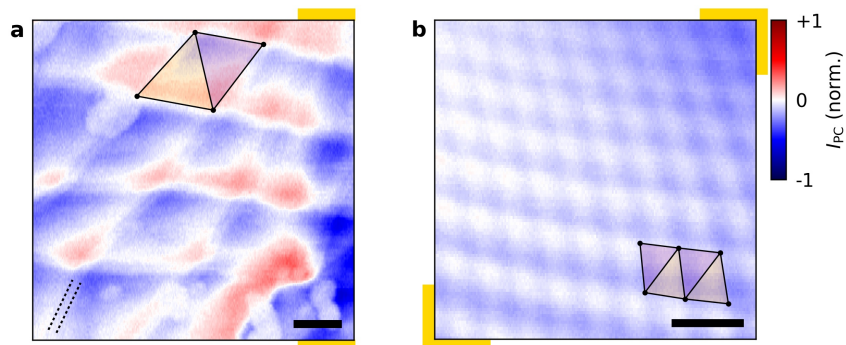

**Figure S2 Near-field photocurrent images in other regions/devices. a** Photocurrent image taken in one of our other devices at $E$ = 117 meV near charge-neutrality. The profile is qualitatively the same as in Fig. 2**a** of the main text, due to a similar arrangement of domain walls with respect to the current path lines between measurement contacts. **b** Photocurrent image of a higher-density network with moiré domains as small as ~100 nm, measured at $E$ = 188 meV near charge-neutrality. In both panels the yellow/purple triangles indicate the location of AB/BA domains, and the gold patches give a rough indication of the source/drain contacts. Length of scale bar is 200 nm in both panels.

## Supplementary Section 3: Photo-thermoelectric effect

Our description of the photo-thermoelectric effect (PTE) is based on two local linear response equations. The first reads

$$\boldsymbol{J}(\boldsymbol{r}) = -\sigma(\boldsymbol{r})\boldsymbol{\nabla}V(\boldsymbol{r}) - \sigma(\boldsymbol{r})S(\boldsymbol{r})\boldsymbol{\nabla}\delta T(\boldsymbol{r}), \tag{S1}$$

where $\boldsymbol{J}(\boldsymbol{r})$ is the electric current density, $\sigma(\boldsymbol{r})$ is the local, direct-current (DC) conductivity, $V(\boldsymbol{r})$ the electric potential, $S(\boldsymbol{r})$ the Seebeck coefficient, and $\delta T(\boldsymbol{r}) \equiv T(\boldsymbol{r}) - T_0$ is the temperature deviation from the substrate temperature $T_0$, which we assume to be constant. The first term is simply the local form of Ohms law, while the second represents the Seebeck effect, i.e. an electric current driven by a temperature gradient.

The second equation involves the heat current density $\boldsymbol{q}(\boldsymbol{r})$ and is given by

$$\boldsymbol{q}(\boldsymbol{r}) = -\kappa(\boldsymbol{r})\boldsymbol{\nabla}\delta T(\boldsymbol{r}) + \Pi(\boldsymbol{r})\boldsymbol{J}(\boldsymbol{r}), \tag{S2}$$

where $\kappa(\boldsymbol{r})$ is the thermal conductivity and $\Pi(\boldsymbol{r}) = T(\boldsymbol{r})S(\boldsymbol{r}) \approx T_0 S(\boldsymbol{r})$ is the Peltier coefficient. The first term describes the normal heat conduction (Fourier law) while the second describes the heat current generated by a flowing electric current, known as the Peltier effect.

At steady state, the following continuity equations for the two currents hold

$$\boldsymbol{\nabla} \cdot \boldsymbol{J}(\boldsymbol{r}) = 0, \tag{S3}$$

$$\boldsymbol{\nabla} \cdot \boldsymbol{q}(\boldsymbol{r}) = -g(\boldsymbol{r})\delta T(\boldsymbol{r}) + Q(\boldsymbol{r}). \tag{S4}$$

The first equation simply expresses charge conservation, while the second represents heat dissipation to the substrate (first term, $g(\boldsymbol{r})$ being the thermal coupling to the substrate) or heat generation by light absorption (second term) as described in the main text.

Taking the divergence of (S2) and using (S3) and (S4) yields the temperature diffusion equation

$$-\boldsymbol{\nabla}[\kappa(\boldsymbol{r})\boldsymbol{\nabla}\delta T(\boldsymbol{r})] + g(\boldsymbol{r})\delta T(\boldsymbol{r}) = Q(\boldsymbol{r}) - T_0 \boldsymbol{\nabla}S(\boldsymbol{r}) \cdot \boldsymbol{J}(\boldsymbol{r}), \tag{S5}$$

where we can distinguish two source terms: the external heat introduced into the system $Q(\boldsymbol{r})$ and the heat generated by a flow of electric current via the Peltier effect. Note that the heat generated through the Joule effect by the steady current $\boldsymbol{J}(\boldsymbol{r})$ is neglected since it is quadratic in $\boldsymbol{J}(\boldsymbol{r})$, and hence quadratic in the absorbed power. If we assume that both $\kappa$ and $g$ are spatially homogeneous, Eq. (S5) simplifies to

$$-\nabla^2 \delta T(\boldsymbol{r}) + L_{\text{cool}}^{-2}\delta T(\boldsymbol{r}) = \frac{1}{\kappa}[Q(\boldsymbol{r}) - T_0 \boldsymbol{\nabla}S(\boldsymbol{r}) \cdot \boldsymbol{J}(\boldsymbol{r})], \tag{S6}$$

where $L_{\text{cool}} \equiv \sqrt{\kappa/g}$ is the cooling length that determines how far the heat can travel in the sample before being lost to the substrate because of out-of-plane conduction.

Equations (S1-4) can be combined into a linear system of Partial Differential Equations (PDEs) in the form

$$-\boldsymbol{\nabla} \cdot \left[\begin{pmatrix} \sigma(\boldsymbol{r}) & T_0\sigma(\boldsymbol{r})S(\boldsymbol{r}) \\ T_0\sigma(\boldsymbol{r})S(\boldsymbol{r}) & T_0^2\sigma(\boldsymbol{r})S^2(\boldsymbol{r}) + T_0\kappa(\boldsymbol{r}) \end{pmatrix}\begin{pmatrix} \boldsymbol{\nabla}V(\boldsymbol{r}) \\ T_0^{-1}\boldsymbol{\nabla}\delta T(\boldsymbol{r}) \end{pmatrix}\right] + \begin{pmatrix} 0 \\ g(\boldsymbol{r})\delta T(\boldsymbol{r}) \end{pmatrix} = \begin{pmatrix} 0 \\ Q(\boldsymbol{r}) \end{pmatrix}. \tag{S7}$$

Here we put the equations in a form that makes explicit the symmetry of the coefficient matrix due to Onsager relations.

At the *m*-th contact the voltage has a constant value $V(\boldsymbol{r}) = V_m$, while the current flowing in it is given as $I_m = \int_{\text{Contact } m} \boldsymbol{J}(\boldsymbol{r}) \cdot \hat{\boldsymbol{n}} \, ds$, $\hat{\boldsymbol{n}}$ being the outward normal unit vector (we consider positive currents those leaving the device).

In our experiment $V_1 = 0$ and $I_2, I_3 = 0$ (see Fig. 3d of the main text for contact numbering). These conditions, together with the boundary conditions on the temperature field $\delta T(\boldsymbol{r}) = 0$ in the contacts, and $\boldsymbol{q}(\boldsymbol{r}) \cdot \hat{\boldsymbol{n}} = 0$ on the rest of the boundary, specify uniquely the solution[5] of the problem (S7) given the heat source $Q(\boldsymbol{r})$.

Solving (S7) numerically via finite element method (FEM) allows calculating $V_2[Q]$ and $V_3[Q]$ from the solution. Because of the linearity of the problem it is in principle possible to solve the system (S7) for a point source located at $\boldsymbol{r}_0$, i.e. $Q_{\text{point}}(\boldsymbol{r}, \boldsymbol{r}_0) = \delta(\boldsymbol{r} - \boldsymbol{r}_0)$ and obtain the results for a generic source in the form

$$V_{2/3}[Q] = \int d\boldsymbol{r}_0 \, \mathcal{R}_{\text{PTE}}^{(2/3)}(\boldsymbol{r}_0) Q(\boldsymbol{r}_0), \tag{S8}$$

where the photovoltage responsivities $\mathcal{R}_{\text{PTE}}^{(m)}(\boldsymbol{r}_0)$ are obtained by evaluating $V_m$ on the solution corresponding to the point source $Q_{\text{point}}(\boldsymbol{r}, \boldsymbol{r}_0)$. This approach is numerically intense, since it requires calculating the solution of (S7) one time for every position at which we want to know the responsivities. We can instead make use of the elegant reciprocity principle[5,6] to solve for the responsivities in one shot.

This reciprocity principle affirms, for our experimental configuration, that $\mathcal{R}_{\text{PTE}}^{(2)}(\boldsymbol{r}_0)$ is equal to the normalized temperature field $\delta T(\boldsymbol{r})/(I_0 T_0)$ obtained by solving (S7) with $V_1 = 0$, $I_2 = I_0$, and $I_3 = 0$, while $\mathcal{R}_{\text{PTE}}^{(3)}(\boldsymbol{r}_0)$ is equal to the temperature field obtained by solving (S7) with $V_1 = 0$, $I_2 = 0$, and $I_3 = I_0$. Solving these two PDE problems with the FEM code[5] we obtained the responsivity maps shown in the main text. We note that this picture is modified in presence of resonant response (either in the sample or the substrate). In this case $Q(\boldsymbol{r}, \boldsymbol{r}_{\text{tip}})$ can spread considerably away from the tip[2] giving rise to additional features.

In the following, we describe the details of the parameters that we feed into the simulations. First, based on sample characterization via AFM/s-SNOM/Raman/photocurrent measurements, we define the sample geometry with specific regions consisting of SLG, and other regions of mTBG (see Fig. 3d of the main text). Next, using the scattering data as a guide (Fig. 1c of the main text), we draw the network of domains in the region of interest, as reflected by Fig. 3e of the main text. Using this geometry, we generate a sample mesh with a variable cell size, with those closest to the domain walls having the smallest edge size of about 1.5 nm.

We define at each cell of the mesh the input parameters as follows:

- In mTBG we consider the domain wall to be of the shear-type[7]. This means that we should take the xx-component of the conductivity and Seebeck tensors[8] as defined in the Supplementary Section 5. We evaluate the Seebeck coefficient $S_{xx}$ and and DC conductivity $\sigma_{xx}(\omega = 0)$ at $T_0 = 300$ K with an energy broadening $\eta = 10$ meV corresponding to a scattering time of $\sim 400$ fs.
- In SLG, we use the Mott formula $S(\mu) = -\frac{\pi^2 k_B^2 T}{3e} \frac{1}{\sigma} \frac{d\sigma}{d\mu}$, with $\sigma(n) \propto 1 + \frac{n(\mu)}{n^*}$ as the DC conductivity, $\mu = \hbar v \sqrt{\pi n}$, and $n^* = 8 \cdot 10^{10}$ cm$^{-2}$ as the impurity density determined for our device – in agreement with what is expected for graphene on hexagonal-boron nitride[9]. To simulate accurately the influence of SLG in the potential landscape, the conductivity serving

- as input for the simulation is set to 1.5x the value of AB-stacked BLG, which is a typical for hBN-encapsulated SLG/BLG devices[10].
- The thermal conductivity $\kappa$ is given at each point using the Wiedemann-Franz law, $\kappa = \frac{\pi^2 k_B^2 T}{3e^2}\sigma$, using the electronic conductivities as defined above.
- The thermal coupling to the substrate is in large part governed via coupling of hot electrons to hBN phonons[11] with a coupling coefficient $g \approx 5 \cdot 10^4$ WK$^{-1}$m$^{-2}$. We take this value for the mTBG and SLG regions, leading to a cooling length $L_{\text{cool}} \approx 270$ nm in mTBG. This compares well with the experimental cooling length extracted in mTBG (see Supplementary Section 7).

## Supplementary Section 4: Photo-thermoelectric effect in a 1D channel

A better grasp of the qualitative features of the responsivities can be obtained by solving the thermoelectric equations in a one-dimensional (1D) setting, where calculations can be carried out analytically. We consider a 1D channel wherein all the variables depend only on the longitudinal ($x$) coordinate. By symmetry, all the vector fields point in the $\hat{x}$ direction. As a further simplification we consider $\sigma, \kappa$ and $g$ as constants, independent on the position. Moreover, we assume that two contacts are placed at $\pm L/2$, with $L \gg L_{cool} = \sqrt{\kappa/g}$, and that heat sources $Q(x)$ and Seebeck coefficient gradients $\partial_x S(x)$ are localized far from the contacts. The continuity equation (S3) can be directly integrated, yielding

$$j_x(x) = \frac{I}{W}, \tag{S9}$$

The potential difference between the two contacts can be obtained by integrating the electric current equation (S1) and reads

$$\Delta V \equiv V\left(-\frac{L}{2}\right) - V\left(\frac{L}{2}\right) = IR + \int_{-\frac{L}{2}}^{\frac{L}{2}} dx' S(x') \partial_x \delta T(x') = IR - \int_{-\frac{L}{2}}^{\frac{L}{2}} dx' \partial_x S(x') \delta T(x'). \tag{S10}$$

Here, $R = L/(W\sigma)$ is the resistance and we integrated by parts the last term imposing the boundary condition $\delta T(x) = 0$ for $x = \pm L/2$. The heat conduction equation (S6) can be inverted using the Green function of the 1D diffusion equation that is decaying far from the origin

$$[-\partial_x^2 + L_{cool}^{-2}] \frac{L_{cool}}{2} e^{-\frac{|x|}{L_{cool}}} = \delta(x), \tag{S11}$$

yielding

$$\delta T(x') = \int_{-L/2}^{L/2} dx'' \frac{L_{cool}}{2} e^{-\frac{|x'-x''|}{L_{cool}}} \frac{1}{\kappa} \left[Q(x'') - \frac{IT_0}{W} \partial_x S(x'')\right]. \tag{S12}$$

In an open-circuit configuration ($I = 0$) the above equations simplify and we can write the photovoltage, substituting (S12) into (S10) as

$$\Delta V = W \int_{-\frac{L}{2}}^{\frac{L}{2}} dx'' \mathcal{R}(x'') Q(x''), \tag{S13}$$

with the photovoltage responsivity given by

$$\mathcal{R}(x'') = -\frac{1}{\kappa W} \int_{-\frac{L}{2}}^{\frac{L}{2}} dx' \frac{L_{cool}}{2} e^{-\frac{|x'-x''|}{L_{cool}}} \partial_x S(x'). \tag{S14}$$

Note that, according to (S12) the normalized temperature profile in absence of heat sources but with $I = I_0$ reads

$$\frac{\delta T(x')}{I_0 T_0} = -\frac{1}{\kappa W} \int_{-\frac{L}{2}}^{\frac{L}{2}} dx'' \frac{L_{cool}}{2} e^{-\frac{|x'-x''|}{L_{cool}}} \partial_x S(x''), \tag{S15}$$

thus explicitly confirming the validity of the reciprocity approach in this simplified 1D setting.

## Supplementary Section 5: Calculation of thermoelectric transport coefficients in mTBG

The strategy to calculate the relevant thermodynamic quantities of mTBG in the vicinity of domain walls, including the DC conductivity tensor $\sigma_{\alpha\beta}(\omega)$ and the Seebeck tensor $S_{\alpha\beta}$, is outlined in three steps. First, we calculate the band structure for different stacking configurations of bilayer graphene ranging from AB-stacking, to the saddle-point (SP) configuration, and to the BA-stacking. Each of these configurations corresponds to a different displacement vector describing the relative lateral displacement between the two graphene layers, as depicted in a Fig. S3 of a shear-type domain wall. Second, once the band structures are known, we calculate $\sigma_{\alpha\beta}(\omega)$ and $S_{\alpha\beta}$ for each of the stacking configurations. Finally, we define the spatial profile of the displacement vector in mTBG, yielding the spatial profile of $\sigma_{\alpha\beta}(\omega)$ and $S_{\alpha\beta}$.

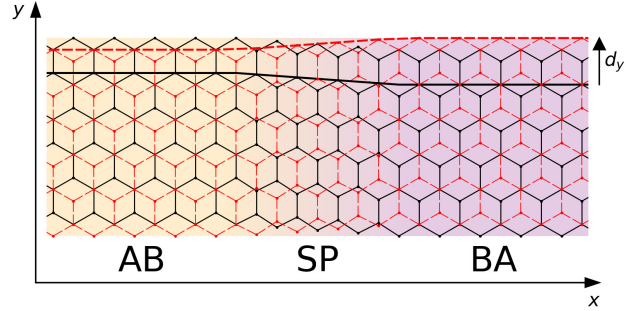

**Figure S3 Local stacking configuration for a shear-type domain wall.** The displacement $d_y$ increases by $a_0$ when translating from an AB region to a BA region, while crossing the saddle-point configuration.

**Band structure for bilayer graphene.** We calculate the electronic band structure of bilayer graphene with an arbitrary displacement between the two layers. Using the model in Ref. 8, we define the displacement vector $d_y \mathbf{e}_y$ along the y-direction. With $d_y$ defined in units of nearest-neighbour distance $a_0 = 0.142$ nm, $d_y = 1, 1.5, 2$ correspond to AB, SP, BA stacking configurations respectively.

For each stacking configuration between the AB and BA sites, we consider an infinite lattice of that configuration, and solve the eigenvalue problem

$$\mathcal{H}(\mathbf{k}) u_\nu(\mathbf{k}) = \epsilon_\nu(\mathbf{k}) u_\nu(\mathbf{k}), \tag{S16}$$

with the effective 4×4 Hamiltonian matrix given by

$$\mathcal{H}(\mathbf{k}) = \begin{bmatrix} H_0^+ & U^\dagger \\ U & H_0^- \end{bmatrix}, \tag{S17}$$

in which $H_0$ is the Hamiltonian of single layer graphene (SLG), and $U$ describes the interaction potential between the two layers

$$H_0^\pm(\mathbf{k}) = \begin{bmatrix} \pm\Delta/2 & \hbar v(\xi k_x + i k_y) \\ \hbar v(\xi k_x - i k_y) & \mp\Delta/2 \end{bmatrix}, \tag{S18}$$

$$U = \frac{\gamma_1}{3}\left(1 + 2\begin{bmatrix} \cos\left(\frac{2\pi}{3} d_y\right) & \cos\left(\frac{2\pi}{3}(d_y+1)\right) \\ \cos\left(\frac{2\pi}{3}(d_y-1)\right) & \cos\left(\frac{2\pi}{3} d_y\right) \end{bmatrix}\right). \tag{S19}$$

Here $\Delta$ is the interlayer potential energy at each of the two layers, $v \approx 1 \times 10^6$ m/s is the band velocity of SLG, $\xi = \pm 1$ selects between the $K$ and $K'$ valley and $\gamma_1 \approx 0.4$ eV is the interlayer coupling.

Since we change the carrier density in mTBG solely with one gate, the chemical potential $\mu$ is always positioned outside of the displacement field induced bandgap, and therefore any corrections to this model due to interface states[12,13] are beyond the scope of this work.

In our model we assume the applied back gate voltage fixes the carrier density $n$ everywhere in our mTBG device, and that the chemical potential $\mu(d_y, n)$ varies spatially due to the varying density of states found for different stacking configurations present in the moire lattice of mTBG. To calculate the chemical potential, we first fix the interlayer potential through a simple capacitance model for bilayer graphene above a backgate. In this model $n(\Delta) = \frac{2\epsilon_0}{e^2 d_0}\Delta$, where $d_0 = 0.34$ nm corresponds to the vacuum distance between two layers. Knowing the density of states for every stacking configuration and calculated interlayer potential, we then build a spatial profile of the spatially varying chemical potential in our devices $\mu(d_y, n)$.

**Calculation of the Seebeck and conductivity tensors.** Following the same definitions as in Ref. 14, we calculate for each stacking configuration the Seebeck tensor $S_{\alpha\beta}$ and the optical conductivity $\sigma_{\alpha\beta}(\omega)$. These tensors relate respectively the heat and current responses in the $\alpha$ direction under an applied electric field in the $\beta$ direction. The Seebeck tensor is defined under the Relaxation Time Approximation as

$$S_{\alpha\beta} = -\frac{1}{eT}\frac{\mathcal{W}^{(1)}_{\alpha\beta}}{\mathcal{W}^{(0)}_{\alpha\beta}}, \tag{S20}$$

with

$$\mathcal{W}^{(p)}_{\alpha\beta} \equiv -\pi g \sum_\nu \int \frac{d^2\boldsymbol{k}}{(2\pi)^2} f'_{\boldsymbol{k}\nu} \cdot (\epsilon_{\boldsymbol{k}\nu} - \mu)^p \langle u_{\boldsymbol{k}\nu}|\partial_{k_\alpha}\widehat{H}(\boldsymbol{k})|u_{\boldsymbol{k}\nu}\rangle \langle u_{\boldsymbol{k}\nu}|\partial_{k_\beta}\widehat{H}(\boldsymbol{k})|u_{\boldsymbol{k}\nu}\rangle. \tag{S21}$$

Here, $g = 4$ corresponds to the fourfold valley/spin degeneracy, $f'$ is the derivative of the Fermi-Dirac distribution $f$. Furthermore, $\nu$ counts over the four bands and $\partial_{k_j}$ is the momentum derivative in the direction $j$. For the electrical conductivity we used the Kubo formula[14] with an energy broadening $\eta = 10$ meV.

**Profile of the displacement vector.** Previous work[7] has experimentally determined the profile of the displacement vector $d_y(x_d)$ for a distance $x_d$ to the middle of a domain wall. This profile can be described as

$$d_y(x_d) = 1 + \frac{2}{\pi}\arctan(e^{\pi x_d/l_{DW}}), \tag{S22}$$

with $l_{DW} = 6.2$ nm encoding the width of a shear domain wall, and $l_{DW} = 10.1$ nm corresponding to the width of the energetically less-favoured tensile domain wall.

By combining the definitions in this section, we obtain a spatial map of $\sigma_{\alpha\beta}(\omega)$ and $S_{\alpha\beta}$ in a network of domain walls, by evaluating at each position the distance $x_d$ to the nearest domain.

**Supplementary Section 6: Impact of strain inside the AB domains**

As mentioned in the main text, the spatially moving zero-crossing observed in the gate-voltage response (dotted line in Fig. 2d of the main text) is not captured in our PTE model. To illustrate this, Fig. S4a shows our simulation of the gate-voltage response presented in the main text and Supplementary Section 3. It plots the simulated photovoltage for various carrier concentrations along the line trace that represents the experimental data of Fig. 2d of the main text. In comparison with the experimental data, our model accounts for two sign changes: one at charge neutrality (CNP, $n$ = 0), and the other away from CNP and depending on the position ($x_{tip}$) within the moiré domain. The sign reversal at CNP is caused by the sign-reversal of the Seebeck coefficient with charge polarity, whereas the spatially dependent sign change away from CNP within the domains is a result of a competition of opposing photoresponse from opposite domain walls. Since the generated photovoltage is antisymmetric with respect to the domain wall (see Fig. 2c of the main text), positions of zero-crossings in the spatial profile of the photovoltage map depend strongly on the alignment of nearby domains to the projection of current flows (Fig. 3f,g of the main text). Despite this complex interplay, the spatial Seebeck profile in our model does not reveal a spatially dependence of the photovoltage sign change away from CNP.

To obtain a better understanding of the possible origins of the spatially dependent sign change, we explore two modifications to the Seebeck coefficient used in our current model and their impact on the calculated photovoltage maps. The first modification consists of a spatially varying Seebeck coefficient in the moiré unit cell, for example, due to strain[15]. Figure. S4b plots an example of a spatially varying Seebeck coefficient, similar to the one presented in the main text, but modelled to be ~1.5% lower in the central part of a domain. This profile essentially describes the case where the AB/BA stacked regions are more strained close to edges of the domains, which seems reasonable considering the large strain distributions found close to domain walls[7,16]. Considering a 1D system, this particular modification will generate a photoresponse that has an antisymmetric profile between two domains (see Eq. (2) of the main text), but with an opposite sign as compared to the photoresponse generated by the domain walls themselves. The second modification we consider is an enhancement of the Seebeck coefficient at the domain walls (see Fig. S4b,c). The reason for this is two-fold. First, we note that our originally calculated Seebeck coefficient at the domain wall (green dashed line in Fig. S4c, corresponding to a SP site) is significantly smaller than that at the middle of the domain (green solid line in Fig. S4c, corresponding to an AB site). Because the difference between Seebeck coefficients at the SP and AB sites remains large for all doping levels in our model, the photoresponse also remains strong away from CNP (Fig. S4a). However, our measured photovoltage instead decays rapidly with increasing doping (Fig. 2e of the main text), suggesting the difference between Seebeck coefficients at the two sites should reduce with increasing carrier density. Second, a reduced photovoltage generation by the domain walls allows any other small photovoltage contribution to yield a measurable influence, for example, one arising from a spatially varying Seebeck coefficient in the AB stacked domains.

We employ a simple toy model to evaluate the effect of these modifications in our mTBG lattice, considering the same measurement configuration as in Fig. 2a of the main text. We approximate the lattice structure as a square lattice of domains, together with a domain running along the diagonal. However, since the current flow direction is parallel to the diagonal domain in this measurement configuration, we can consider its contribution to the measured photovoltage neglectable (Fig. 2a of the main text). Whilst the diagonal domain should in principle influence the strain profile, we neglect this additional complication to gain some intuition into the possible effects of strain. Therefore, we consider the spatial photoresponse for a 1D channel (Eq. (2) of the main text) that crosses two parallel domains. Since the current flow direction is aligned to the diagonal, both the horizontal and vertical domains contribute equally to the total photovoltage and thus we can simply sum up both contributions to build a spatial map.

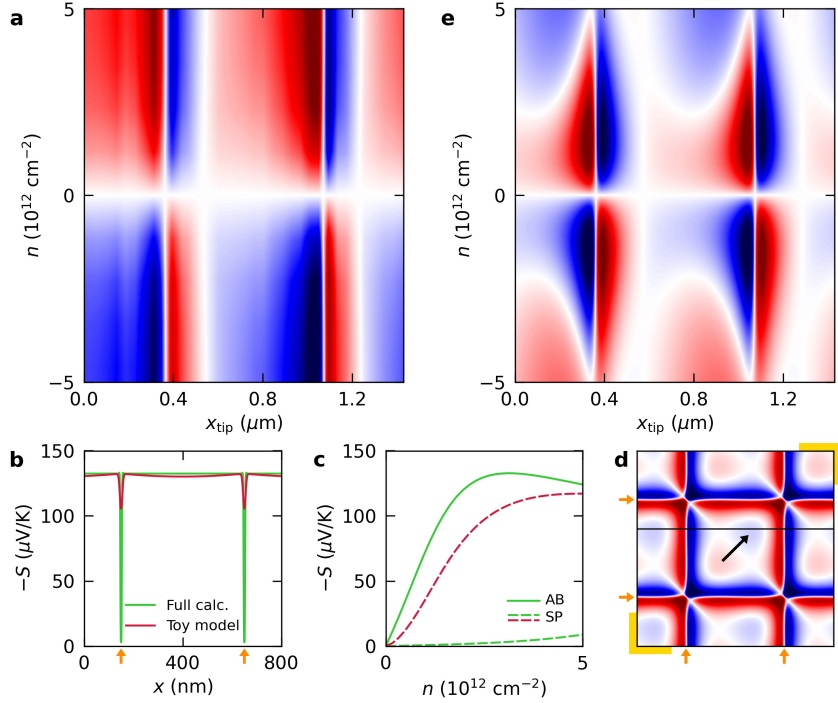

**Figure S4 Additional contribution to the Seebeck profile due to strain. a** Plot of the simulated photovoltage along the same line cut as in the experimental data (Fig. 2**d** of the main text) for a range of carrier densities ($n$). The simulated $V_{PV}$ is calculated using the same method as presented in Supplementary Section 3. Length of scale bar: 500 nm; colour code: blue: -1, red: +1. **b** Calculated Seebeck coefficient according to the full calculation (green) crossing two domain walls at $n = 3 \times 10^{12}$ cm$^{-2}$, serving as input for the photovoltage simulation shown in panel **a**. Our toy model (red) uses an enhanced Seebeck coefficient at the domain walls (position indicated by the orange arrows) and a slightly reduced Seebeck coefficient between the domains. **c** Carrier density dependence of the Seebeck coefficient at the AB and SP sites (corresponding to the middle of the domain and domain wall, respectively) for both models, highlighting the enhanced Seebeck coefficient at the SP in our toy model. **d** Spatial profile of the calculated photovoltage in our toy model with $n = 3 \times 10^{12}$ cm$^{-2}$ with the position of domain walls indicated by the orange arrows. The black arrow marks the current flow direction, with the position of the contacts indicated by the gold corners. **e** Line trace of the calculated photovoltage based on our toy model as function of carrier density. The line trace is taken along the black line in panel **d**.

Figure S4d shows such a modelled spatial map for a fixed $n = 3 \times 10^{12}$ cm$^{-2}$, which highlights the interplay of various photovoltage contributions. In particular, two patches of photoresponse emerge with opposite sign inside the moiré domains, and change polarity at CNP. Fig. S4e plots the gate-voltage response calculated in our toy model, taken across the line trace drawn in Fig. S4d. Importantly, it contains the main features present in our full calculation (Fig. S4a): sign reversals of the photoresponse at CNP and spatially in the domains. In addition, the details show some closer resemblance to our experimental data (Fig. 2d of the main text). In this toy model, the photovoltage from the domain walls reduces towards higher carrier densities, allowing the contribution due to the spatially-varying Seebeck coefficient to dominate. The competition between these two contributions causes the sign-change (away from CNP) to vary spatially as observed in our experimental data (Fig. 2d of the main text). This simple toy model illustrates qualitatively the possible influence of strain and enhanced Seebeck coefficients at the domain walls in influencing the photoresponse of our mTBG devices, even though the exact two-dimensional strain profile, including the influence of the diagonal domains has been ignored.

## Supplementary Section 7: Cooling length in our devices

The photoresponse generated by the photo thermoelectric effect is driven by local temperature gradients generated in the electron gas in the vicinity of inhomogeneities in the Seebeck coefficient, which decay over a characteristic length scale from the source. Microscopically, this corresponds to the distance over which initial photoexcited carriers equilibrate with the lattice and is referred to as the cooling length $L_{cool}$. In the PTE, a photocurrent can be generated as long as photoexcitation occurs within a typical distance $L_{cool}$ from any junction that exhibits gradients in the Seebeck coefficient. In Bernal stacked bilayer graphene, the cooling length has been measured to be around ∼250 nm[17]. This is why in our mTBG samples the spatial photocurrent profile is so complex, because thermal gradients generated by photoexcitation in the middle of moiré domains can reach different surrounding junctions, which add linearly and contribute to the globally measured photoresponse. As mentioned in the main text, our samples tend to have extrinsic junctions in the form of stacking faults that also generate photocurrent and, in the case that they are located a distance $L_{cool}$ from the superlattice region, would contribute a background signal to the photoresponse measured in our moiré domains. They could explain, for example, the constant negative photoresponse observed in the moiré domains at high doping levels (Fig. 2d and Fig. 4a of the main text). To rule out such contributions and allow correct interpretation of the photoresponse from moiré domains alone, we studied how the photoresponse from stacking faults behaves and measure $L_{cool}$ in our devices.

Fig. S5a plots an extended photocurrent map of the device presented in the main text (Fig. 4a of the main text) at a carrier density $n \sim 4 \times 10^{12}$ cm$^{-2}$, that includes the single layer graphene (SLG) region. The map clearly shows photocurrent hot spots on one side of the device originating from SLG-mTBG interface (marked by green dotted line), and on the other side from cracks/stacking faults (black dotted line). In between these interfaces we observe the anomalous negative photoresponse in the moiré domains. However, the photocurrent hotspots can be seen to decay around 1 µm into the sample, suggesting another possible origin to the photoresponse observed in the moiré domains. To evaluate this, we extracted $L_{cool}$ from the SLG-mTBG interface. Figure. S5b plots a line trace taken across the SLG-mTBG interface that extends a few microns into the mTBG region. Since the photocurrent profile is locally invariant under translations along the SLG-mTBG interface, we use the 1D version of our model of the PTE (Eq. (2) of the main text). This simplifies the analysis, as the photocurrent on either side of the interface is simply proportional to increase in electron temperature $\delta T$ (ignoring the domain walls further away) caused by local heating. That said, we should consider that heat spreads radially away from the excitation position, according to the following profile[18]

$$\delta T(x) \propto K_0\left(\sqrt{\left((x - x_\text{tip})^2 + L_\text{tip}^2\right)/L_\text{cool}^2}\right), \tag{S23}$$

where $K_0$ is the modified Bessel function of the second kind, $L_\text{tip}$ corrects for the finite radius of the AFM tip, and $L_\text{cool}$ is the cooling length.

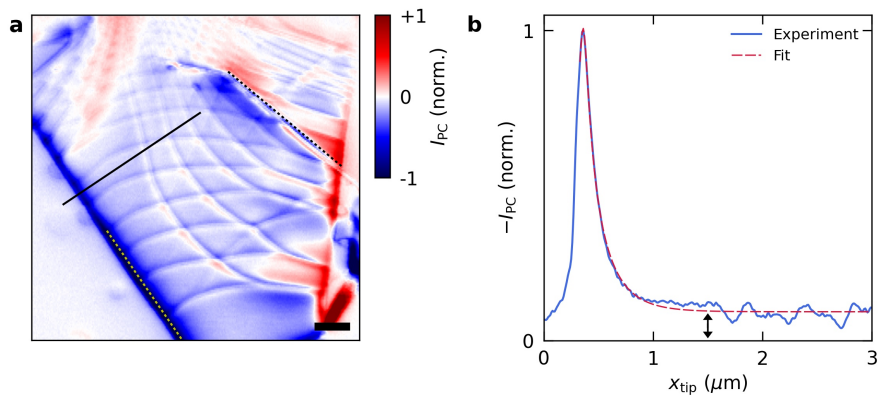

**Figure S5 Extraction of cooling length. a** Extended photocurrent map of Fig. 4a of the main text ($E$ = 117 meV, $n \sim 4 \times 10^{12}$ cm$^{-2}$). The yellow dotted line marks partially marks the interface between SLG on the left side and mTBG on the right side, while the black dotted line marks a crack/stacking fault in our device. Length of scale bar: 500 nm. **b** Line trace of the photocurrent taken along the black line in panel **a**. The peak in photocurrent marks the position of the SLG-mTBG interface. A fit of the photocurrent profile yields a cooling length of 240 nm, and a background offset whose magnitude is a fraction ∼0.1 of the photocurrent generated from the SLG-mTBG interface.

Importantly, we find Eq. (S23) alone does not describe our experimental data (Fig. S5b). Instead, a constant offset is needed in our fit of Eq. (S23) (see arrow in Fig. S5b) to describe the decay of photocurrent from the SLG-mTBG interface. From the decay of the photocurrent we extract $L_\text{cool}$ = 240 nm. This behaviour, in addition to the constant off-set needed to describe out experimental data, shows that photocurrent generation from the SLG-mTBG is not responsible for the anomalous photocurrent observed at high doping in our devices (Fig. 4a of the main text).

**Supplementary Section 8**: **Effects beyond the photo-thermoelectric effect**

All our experimental evidence points towards an explanation of the observed photoresponse data based on the PTE. Indeed, the FEM simulations that include only this effect are in good agreement with experimental data and small discrepancies can be explained by a more complicated Seebeck coefficient profile as suggested by the simplified model in Supplementary Section 6. However, even if the general picture is well described by PTE, we cannot completely exclude that other effects contribute small additional corrections.

Here, we comment briefly on two other possible mechanisms of photoresponse generation, namely the photogalvanic effect and the photovoltaic effect. The photogalvanic effect generates a photocurrent thanks to the intrinsic *second-order* response of the material to the incident electric field that gives rise to a DC current density $\boldsymbol{j}_{\text{PG}}(\boldsymbol{r},\boldsymbol{r}_{\text{tip}}) \propto |\boldsymbol{E}(\boldsymbol{r},\boldsymbol{r}_{\text{tip}},\omega_{\text{ph}})|^2$. Such a response is symmetry-forbidden for homogeneous materials (for unpolarised light) but can play a role in the presence of strong electronic density gradients or strain.

This modifies our equation (S1) into

$$\boldsymbol{J}(\boldsymbol{r}) = -\sigma(\boldsymbol{r})\boldsymbol{\nabla}V(\boldsymbol{r}) - \sigma(\boldsymbol{r})S(\boldsymbol{r})\boldsymbol{\nabla}\delta T(\boldsymbol{r}) + \boldsymbol{j}_{\text{PG}}(\boldsymbol{r},\boldsymbol{r}_{\text{tip}}), \tag{S24}$$

adding another source term.

We can write a relation similar to Eq. (1) of the main text for the PGE, that reads

$$V_{\text{PGE}}^{(m)} = \int d\boldsymbol{r}\, \boldsymbol{\mathcal{R}}_{\text{PGE}}^{(m)}(\boldsymbol{r}) \cdot \boldsymbol{j}_{\text{PG}}(\boldsymbol{r},\boldsymbol{r}_{\text{tip}}). \tag{S25}$$

Again, $\boldsymbol{\mathcal{R}}_{\text{PGE}}^{(m)}(\boldsymbol{r})$ can be calculated by solving the thermoelectric transport equations using FEM and taking advantage of the Shockley-Ramo theorem[5,6,19], $\boldsymbol{\mathcal{R}}_{\text{PGE}}^{(m)}(\boldsymbol{r})$ is in fact proportional to the gradient of the potential that would be present in the system in absence of sources and biasing the *m*-th with a constant current. Qualitatively, $\boldsymbol{\mathcal{R}}_{\text{PGE}}^{(m)}(\boldsymbol{r})$ is a smooth vector field flowing mainly in the direction connecting the contacts at which photocurrent is measured. Importantly it is very weakly affected by the cooling length. As a consequence, we expect PG features (if any) to be sharper with respect to PTE features that are smoothed on the length scale $L_{\text{cool}}$.

We note also that the photovoltaic effect can also play a role in the photoresponse of systems with more than one energy band. Including this effect in our model would require studying a system of three coupled equations including the imbalance density (electron density + holes density) and current. While this is outside the scope of this work, we tend to exclude this explanation since photovoltaic contributions should display a threshold behaviour as a function of the photon energy that is not observed in our experimental data.